\documentclass[conference]{IEEEtran}
\IEEEoverridecommandlockouts
\usepackage{cite}
\usepackage{amsmath,amssymb,amsfonts}
\usepackage{algorithmic}
\usepackage{graphicx}
\usepackage{algorithm}
\usepackage{algorithmic}
\usepackage{subcaption}
\usepackage{hyperref}

\usepackage{url}
\usepackage{textcomp}
\usepackage{xcolor}
\usepackage{todonotes}
\def\BibTeX{{\rm B\kern-.05em{\sc i\kern-.025em b}\kern-.08em
    T\kern-.1667em\lower.7ex\hbox{E}\kern-.125emX}}
\begin{document}

\title{Anatomy-Preserving Latent Diffusion for Generation of Brain Segmentation Masks with Ischemic Infarct}

\author{
\IEEEauthorblockN{1\textsuperscript{st} Lucia Borrego}
\IEEEauthorblockA{\textit{Department of Diagnostic Imaging} \\
\textit{Hospital de la Santa Creu i Sant Pau}\\
Barcelona, Spain\\
lborrego@santpau.cat}
\and
\IEEEauthorblockN{2\textsuperscript{nd} Vajira Thambawita}
\IEEEauthorblockA{\textit{SimulaMet} \\
Oslo, Norway \\
vajira@simula.no}
\and
\IEEEauthorblockN{3\textsuperscript{rd} Marco Ciuffreda}
\IEEEauthorblockA{\textit{Department of Diagnostic Imaging} \\
\textit{Hospital de la Santa Creu i Sant Pau}\\
Barcelona, Spain \\
mciuffreda@santpau.cat}
\and
\IEEEauthorblockN{4\textsuperscript{th} Inés del Val}
\IEEEauthorblockA{\textit{Department of Diagnostic Imaging} \\
\textit{Hospital de la Santa Creu i Sant Pau}\\
Barcelona, Spain \\
idelval@santpau.cat}
\and
\IEEEauthorblockN{5\textsuperscript{th} Alejandro Dominguez}
\IEEEauthorblockA{\textit{Department of Medical Physics} \\
\textit{Hospital de la Santa Creu i Sant Pau}\\
Barcelona, Spain \\
adominguezp@santpau.cat}
\and
\IEEEauthorblockN{\centering 6\textsuperscript{th} \centering Josep Munuera}
\IEEEauthorblockA{
\textit{\centering Department of Diagnostic Imaging} \\
\textit{\centering Hospital de la Santa Creu i Sant Pau}\\
Barcelona, Spain\\
jmunuera@santpau.cat}
}

\maketitle
\begin{abstract}
The scarcity of high-quality segmentation masks remains a major bottleneck for medical image analysis, particularly in non-contrast CT (NCCT) neuroimaging, where manual annotation is costly and variable. 
To address this limitation, we propose an anatomy-preserving generative framework for the unconditional synthesis of multi-class brain segmentation masks, including ischemic infarcts.

The proposed approach combines a variational autoencoder trained exclusively on segmentation masks to learn an anatomical latent representation, with a diffusion model operating in this latent space to generate new samples from pure noise. 
At inference, synthetic masks are obtained by decoding denoised latent vectors through the frozen VAE decoder, with optional coarse control over lesion presence via a binary prompt.

Qualitative results show that the generated masks preserve global brain anatomy, discrete tissue semantics, and realistic variability, while avoiding the structural artifacts commonly observed in pixel-space generative models. 
Overall, the proposed framework offers a simple and scalable solution for anatomy-aware mask generation in data-scarce medical imaging scenarios.

\end{abstract}

\begin{IEEEkeywords}
Latent diffusion models, anatomy-aware generative modeling, brain segmentation masks, ischemic infarct, non-contrast CT (NCCT), medical image synthesis
\end{IEEEkeywords}
\section{Introduction}
Deep learning has become the dominant paradigm in medical image analysis, achieving state-of-the-art performance in tasks such as segmentation, detection, and disease characterization. However, supervised methods critically depend on the availability of large, diverse, and well-annotated datasets. In neuroimaging, and particularly in non-contrast computed tomography (NCCT), acquiring high-quality pixel-level annotations is especially challenging due to the need for expert neuroradiological knowledge, the time-consuming nature of manual annotation, and inter-observer variability. In addition, privacy regulations and institutional constraints often limit data sharing across centers, further exacerbating data scarcity \cite{b0}.

To address these limitations, synthetic data generation has emerged as a promising strategy for dataset augmentation and improved generalization. Early approaches predominantly relied on Generative Adversarial Networks (GANs), which have been widely applied to medical image synthesis and segmentation-related tasks, including mask-conditioned image synthesis and joint image–mask generation. Despite their success, GAN-based methods suffer from well-known limitations such as training instability, mode collapse, and limited diversity, particularly when modeling complex anatomical structures and multi-class spatial relationships \cite{b01}.

Beyond image synthesis, several works have explored the direct generation of segmentation masks to model anatomical variability, including shape-based statistical models, variational autoencoders (VAEs), and GAN-based mask generators. While these approaches can capture coarse anatomical structure, they often struggle to preserve fine-grained spatial coherence and realistic inter-class relationships, especially in high-resolution, multi-tissue brain segmentation tasks \cite{b2}. Moreover, many methods rely on paired image–mask data or strong shape assumptions, which limits their scalability.

Diffusion probabilistic models (DPMs) have recently emerged as a more stable and expressive alternative to GANs, demonstrating superior sample quality and diversity in both natural and medical image synthesis. Recent studies have applied diffusion models to medical image generation, conditional synthesis, and joint image–mask generation \cite{b1,b2}. However, only a limited number of works have explored the unconditional generation of segmentation masks using diffusion models. These approaches typically operate in pixel space and often produce anatomically implausible results, such as fragmented regions or broken topology, particularly in multi-class settings \cite{b12,b3,b7}.

In this work, we address the underexplored problem of unconditional generation of anatomically coherent brain tissue segmentation masks for NCCT. Instead of performing diffusion in pixel space, we propose an anatomy-preserving latent diffusion framework in which diffusion is carried out in the latent space of a VAE trained exclusively on segmentation masks. The VAE learns a compact latent manifold that encodes global anatomical structure and inter-tissue relationships, acting as a strong anatomical prior. A diffusion model is then trained in this latent space to generate diverse and realistic segmentation masks from pure Gaussian noise \cite{b4}.

By decoupling anatomical structure learning from stochastic generation, the proposed approach reduces the structural artifacts commonly observed in pixel-space diffusion models and eliminates the need for conditioning inputs or paired image–mask data at inference time. As a downstream application, the generated masks can be used as structural inputs for a SPADE-based \footnote{\url{https://github.com/NVlabs/SPADE}} image synthesis framework to generate corresponding NCCT images. Nevertheless, the primary focus of this work is the mask generation pipeline itself.

In summary, this work introduces a simple and effective framework for the unconditional synthesis of anatomically coherent brain segmentation masks. By combining an explicit anatomical latent prior with latent diffusion, the proposed approach bridges classical shape modeling and modern diffusion-based generation, making it suitable for scalable dataset augmentation in annotation-scarce medical imaging scenarios.

The main contributions of this work are summarized as follows:
\begin{itemize}
    \item We propose an anatomy-preserving latent diffusion framework for the unconditional generation of multi-class brain segmentation masks from non-contrast CT data, including ischemic infarct regions, and illustrate its generative capabilities through an interactive model demonstration.
    \item We introduce a mask-only variational autoencoder that learns an explicit anatomical latent space, constraining the generation process to anatomically plausible configurations without requiring conditioning images at inference time.
    \item We demonstrate stable and scalable mask synthesis and release a publicly available synthetic dataset of 605 multi-class segmentation masks, together with pre-trained models, generation scripts, and an online demonstration space to support reproducibility and data augmentation in annotation-scarce settings.
\end{itemize}

\section{Ease of Use}

\subsection{Dataset and Preprocessing}
The dataset used in this study consists of non-contrast computed tomography (NCCT) brain scans acquired from 86 patients diagnosed with ischemic stroke of cardioembolic origin, according to the TOAST (Trial of ORG 10172 in Acute Stroke Treatment) classification criteria. Restricting the cohort to a single etiological subtype was a deliberate choice aimed at reducing heterogeneity in lesion distribution, vascular territory involvement, and underlying pathophysiological mechanisms.

Only NCCT scans acquired during the post-acute phase, defined as 24 to 48 hours after symptom onset, were included. This temporal window was selected for several reasons. First, hyperacute NCCT scans often exhibit subtle or absent parenchymal changes, leading to increased variability and uncertainty in tissue delineation. Second, scans acquired beyond the early post-acute phase may present secondary effects such as edema resorption or structural deformation, which introduce additional variability unrelated to the primary tissue composition. 

All scans were collected retrospectively as part of routine clinical care at Hospital de la Santa Creu i Sant Pau (Barcelona, Spain). Data were acquired using standard clinical NCCT scanners and were fully anonymized prior to processing, in accordance with institutional review board approval and local ethical guidelines.

To ensure anatomical consistency and exclude irrelevant structures, a standardized slice selection strategy was applied to all NCCT volumes. Axial slices were retained only within a restricted cranial range corresponding to 20\%–95\% of the total cranial height. This strategy was designed to systematically remove slices dominated by non-brain anatomy, including:
\begin{itemize}
    \item Cervical and upper neck structures.
    \item The skull base and posterior fossa.
    \item Slices containing large proportions of extracranial tissue or severe imaging artifacts.
\end{itemize}

\begin{figure}[t]
    \centering
    \begin{subfigure}{0.32\linewidth}
        \centering
        \includegraphics[width=\linewidth]{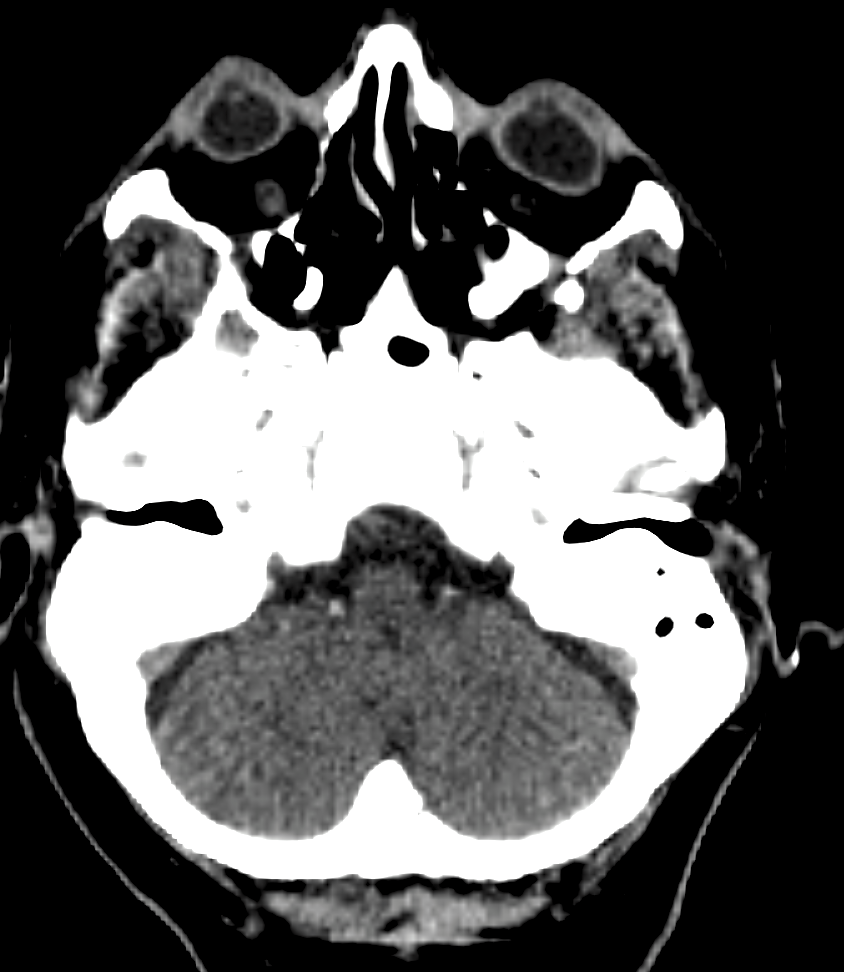}
        \caption{Inferior cervical slice (discarded).}
    \end{subfigure}
    \hfill
    \begin{subfigure}{0.32\linewidth}
        \centering
        \includegraphics[width=\linewidth]{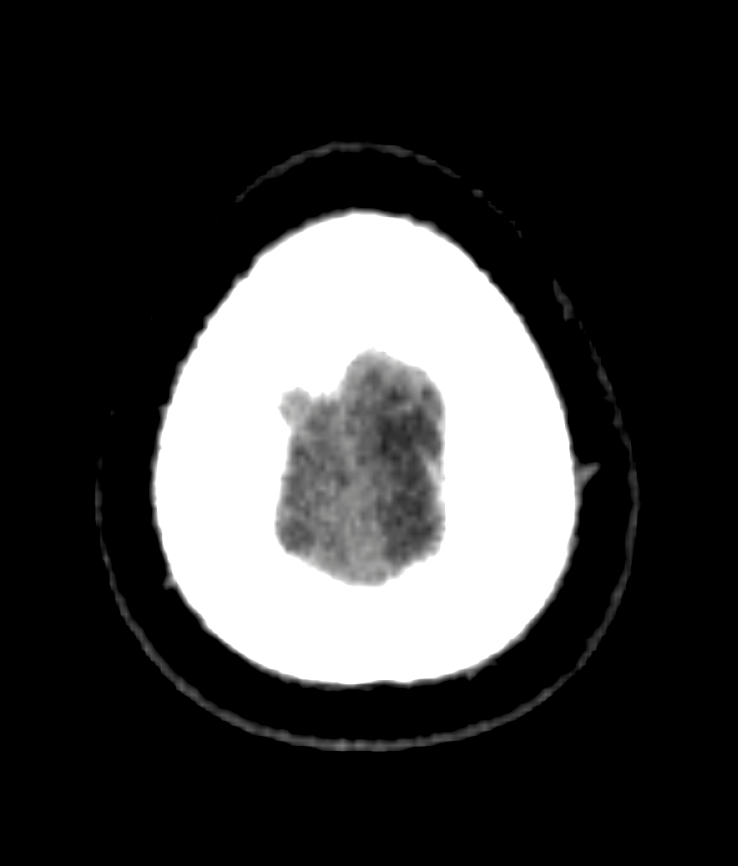}
        \caption{Skull base (discarded).}
    \end{subfigure}
    \hfill
    \begin{subfigure}{0.32\linewidth}
        \centering
        \includegraphics[width=\linewidth]{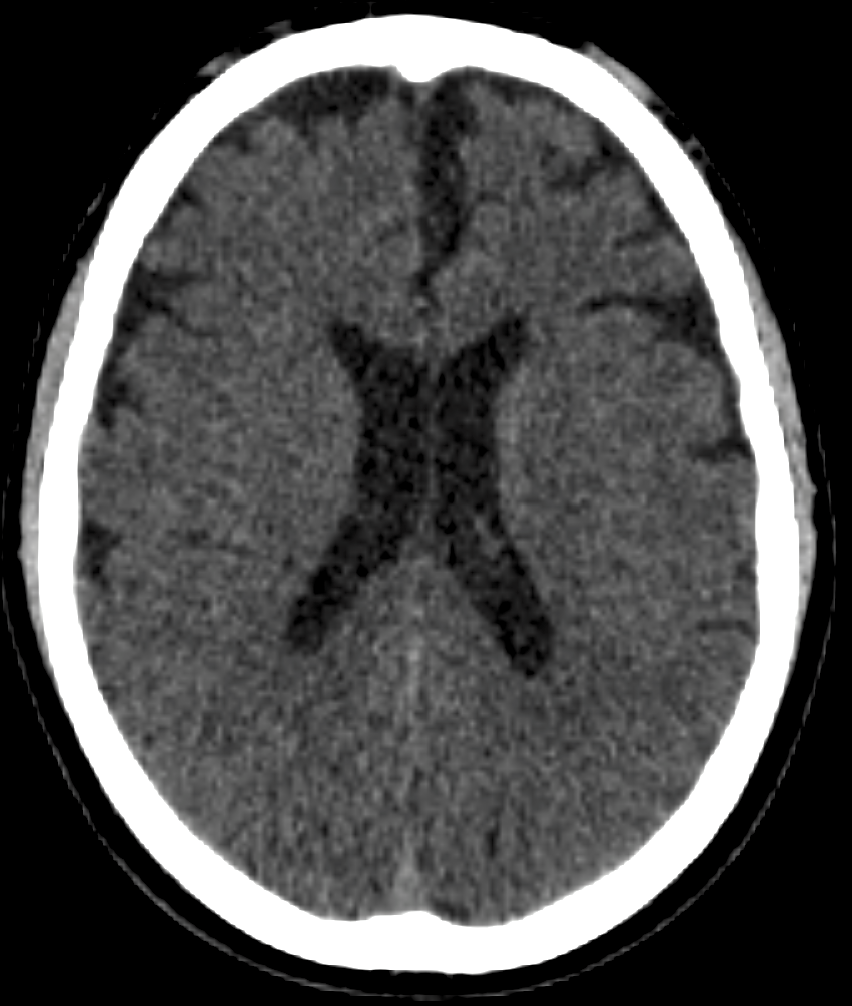}
        \caption{Supratentorial brain slice (retained).}
    \end{subfigure}
    \caption{Illustration of the slice selection strategy used in this study.}
    \label{fig:slice_selection}
\end{figure}

By discarding the inferior and extreme superior portions of the head, the resulting dataset focuses exclusively on supratentorial brain anatomy, which is most relevant for ischemic stroke analysis and downstream modeling tasks. This selection also reduces anatomical variability unrelated to cerebral tissue organization, facilitating more stable and consistent learning of structural patterns by the generative model.

Following patient-level and slice-level selection, a total of approximately 6,500 axial slices were retained across the 86 patients. 

\subsection{Annotation Protocol and Multi-class Mask Construction}
Ischemic infarct regions were manually segmented on axial NCCT slices by two final-year radiology residents with dedicated training in neuroradiology. The resulting lesion annotations were subsequently reviewed, corrected, and finalized by two senior board-certified neuroradiologists. Disagreements were resolved by consensus, yielding clinically reliable infarct ground truth masks.

\begin{figure}
    \centering
    \includegraphics[width=0.75\linewidth]{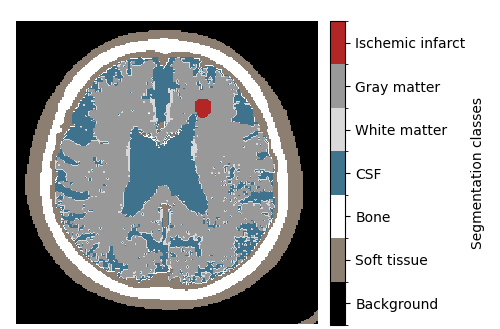}
    \caption{Example of multi-class tissue segmentation on a representative NCCT slice. CTseg was used to automatically segment normal tissue classes, whereas ischemic infarct regions correspond to manual annotations provided by expert annotators.}
    \label{fig:labels}
\end{figure}

To obtain non-lesion tissue labels, we applied an automatic brain tissue segmentation pipeline based on CTseg\footnote{\url{https://github.com/WCHN/CTseg}}, a deep learning framework for multi-class tissue segmentation in non-contrast CT. In brief, CTseg performs dense pixel-wise classification using a fully convolutional neural network that extracts hierarchical features from CT slices and produces anatomically consistent maps of major tissue compartments \cite{b6}. We used CTseg to segment background, cerebrospinal fluid (CSF), skull/bone, and brain parenchyma-related tissues, as show in Figure \ref{fig:labels}. These outputs served as the baseline tissue segmentation for each slice.

Final multi-class masks were created by merging the infarct annotations with the CTseg tissue maps. In slices where infarct pixels overlapped with automatically segmented tissue labels, the infarct label was assigned precedence to ensure that pathological regions were not overwritten by normal tissue classes. This fusion procedure produced a single discrete label map per slice with $C=7$ classes, where one class corresponds explicitly to ischemic infarct.
Across the full cohort of 86 patients, 38 (44.2\%) presented at least one ischemic lesion (acute and/or chronic), while 48 (55.8\%) showed no visible lesions. 
Among lesion-positive cases, 22 patients (25.6\%) exhibited acute lesions only, 9 (10.5\%) presented chronic lesions only, and 7 (8.1\%) showed both acute and chronic lesions. 
For mask generation, acute and chronic infarcts were merged into a single ischemic infarct label, without temporal differentiation. 
Overall, 13 patients (15.1\%) presented more than one lesion, either due to multiple foci of the same type or the coexistence of acute and chronic lesions.

\subsection{Preprocessing and Encoding}
All masks were resized to a fixed spatial resolution of $256 \times 256$ pixels using nearest-neighbor interpolation, preserving discrete boundaries and preventing label mixing. Each mask was converted from a single-channel label map into a one-hot tensor $X \in \{0,1\}^{C \times H \times W}$ to enable categorical reconstruction losses during training. Color palettes were used exclusively for the visualization of qualitative results (Figure \ref{fig:labels}; all training was performed in label space.

\subsection{Data Split and Training Usage}
Data were split at the patient level to prevent slice-level leakage across subjects. A total of 61 patients were assigned to the training set, 8 to the validation set, and 7 to the test set.

At the slice level, the training, validation, and test sets comprised 5,238, 660, and 609 axial slices, respectively. The training set was used to fit the generative models, the validation set to monitor training stability and select checkpoints, and the test set was held out for qualitative evaluation only.

No downstream segmentation networks were trained in this work; the focus is exclusively on unconditional brain mask synthesis.

\section{Proposed Method}

\subsection{Design Rationale}
The proposed framework is guided by a set of deliberate design choices aimed at enabling stable and anatomically coherent unconditional generation of multi-class brain segmentation masks.

First, we avoid direct pixel-space generation of segmentation masks. Multi-class anatomical masks contain sharp boundaries, discrete semantics, and strong topological constraints that are difficult to model directly using pixel-level generative models. In preliminary experiments and prior work, pixel-space diffusion often leads to fragmented regions, label mixing, and anatomically implausible configurations, particularly for rare pathological classes.

Second, we introduce a variational autoencoder trained exclusively on segmentation masks to explicitly learn an anatomical latent manifold. By enforcing categorical reconstruction and latent regularization, the VAE decoder implicitly defines the space of anatomically valid configurations. This design ensures that any sample decoded from the latent space preserves global brain structure, tissue adjacency relationships, and class semantics.

Finally, diffusion is performed in the learned latent space rather than in pixel space. This choice enables efficient sampling with significantly reduced computational cost, while constraining the generative process to remain within the anatomical prior learned by the VAE. The denoising network operates on compact latent vectors and does not need to model spatial structure explicitly, as long-range anatomical dependencies are already encoded in the latent representation.

Together, these design choices decouple anatomical structure learning from stochastic generation, yielding a simple yet robust framework for unconditional mask synthesis.

Let $x \in \{0,\dots,C-1\}^{H \times W}$ be a discrete multi-class segmentation mask with $C=7$ classes and fixed spatial resolution $H=W=256$. We denote its one-hot representation as $X \in \{0,1\}^{C \times H \times W}$. The goal is to synthesize new masks $\tilde{x}$ by sampling from a learned distribution, without using any conditioning CT image at inference time.

\begin{figure}[!t]
    \centering
    \includegraphics[width=1\linewidth]{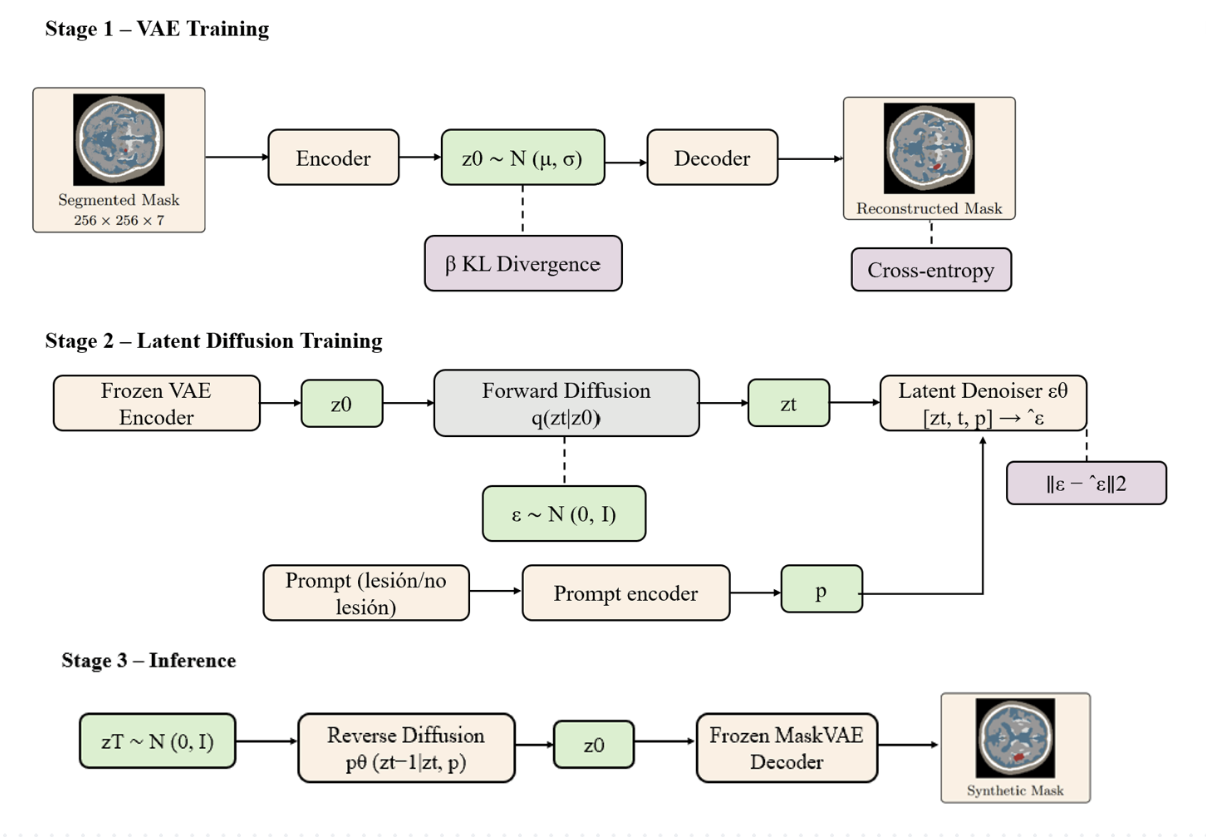}
 \caption{Overview of the proposed anatomy-preserving latent diffusion framework for synthetic brain mask generation using a frozen MaskVAE decoder and latent diffusion.}

    \label{fig:pipeline}
\end{figure}

To introduce minimal controllability over pathology occurrence, each training mask is assigned a binary variable $y \in \{0,1\}$ indicating whether the slice contains any infarct pixels (lesion absent/present). We learn a prompt-conditioned generative model that enables sampling from $p_\theta(x \mid y)$ while keeping the generative process otherwise unconditional (i.e., no spatial guidance).
An overview of the proposed two-stage anatomy-preserving latent diffusion framework is shown in Fig.~\ref{fig:pipeline}. 

\subsection{Stage I: MaskVAE as an Anatomical Prior}
Directly generating discrete multi-class masks in pixel space is challenging because anatomical masks contain sharp boundaries, topology constraints, and strong long-range dependencies (e.g., skull enclosing brain, CSF surrounding parenchyma). To constrain generation to anatomically plausible configurations, we first learn an explicit anatomical prior via a variational autoencoder (VAE) trained exclusively on segmentation masks.

\subsubsection{Architecture}
The MaskVAE encoder $E_\phi$ maps one-hot masks $X$ to the parameters of a diagonal Gaussian posterior distribution:
\begin{equation}
(\mu, \log\sigma^2) = E_\phi(X),
\end{equation}
where $\mu,\sigma \in \mathbb{R}^{D}$ and $D=256$ is the latent dimension. The encoder consists of four strided convolution blocks with kernel size $4$, stride $2$ and ReLU activations, producing a compact feature representation of size $256 \times 16 \times 16$, which is flattened and projected into $(\mu,\log\sigma^2)$.

A latent vector is sampled using the reparameterization trick:
\begin{equation}
z = \mu + \sigma \odot \epsilon, \quad \epsilon \sim \mathcal{N}(0, I).
\end{equation}
The decoder $G_\phi$ maps $z$ back to mask logits $\hat{L} \in \mathbb{R}^{C \times H \times W}$:
\begin{equation}
\hat{L} = G_\phi(z).
\end{equation}
The final discrete reconstruction is obtained by $\hat{x}=\arg\max_c \hat{L}_c$.

\subsubsection{Training Objective}
Reconstruction is treated as pixel-wise multi-class classification. Compared to regression losses (e.g., MSE), a categorical cross-entropy loss preserves discrete semantics and avoids blurred or mixed labels:
\begin{equation}
\mathcal{L}_{\text{rec}} = \mathrm{CE}(\hat{L}, x).
\end{equation}
To regularize the latent distribution toward a standard normal prior and encourage a smooth latent manifold, we include a Kullback--Leibler divergence term:
\begin{equation}
\mathcal{L}_{\text{KL}} = D_{\mathrm{KL}}\left(q_\phi(z\mid X) \,\|\, \mathcal{N}(0,I)\right).
\end{equation}
The final VAE objective is:
\begin{equation}
\mathcal{L}_{\text{VAE}} = \mathcal{L}_{\text{rec}} + \beta \mathcal{L}_{\text{KL}},
\end{equation}
where $\beta$ controls the strength of regularization ($\beta=0.01$ in our implementation).

\subsubsection{Lesion-aware Sampling for Imbalanced Data}
Infarct pixels are rare compared to normal tissue and background. To ensure that the VAE latent space captures pathological configurations, we apply a weighted sampling strategy during VAE training. Each slice receives a sampling weight $w_i$ defined as:
\begin{equation}
w_i =
\begin{cases}
5, & \text{if } \exists (u,v) \text{ such that } x_i(u,v)=\text{lesion},\\
1, & \text{otherwise}.
\end{cases}
\end{equation}
A weighted random sampler draws training batches with replacement according to $\{w_i\}$, increasing the frequency of lesion-containing masks while keeping the loss function unchanged.

\subsection{Stage II: Promptable Latent Diffusion}
After training, the VAE is frozen and used as a fixed anatomical decoder. We then learn a diffusion model in the latent space to sample novel latent codes that remain compatible with the anatomical manifold learned by the VAE.

\subsubsection{Forward Diffusion Process in Latent Space}
For each training mask, we sample $z_0 \sim q_\phi(z\mid X)$. We then draw a timestep $t \sim \mathcal{U}\{0,\dots,T-1\}$ and generate a noisy latent vector:
\begin{equation}
z_t = \sqrt{\bar{\alpha}_t}\, z_0 + \sqrt{1-\bar{\alpha}_t}\,\epsilon,
\quad \epsilon \sim \mathcal{N}(0,I),
\end{equation}
where $\bar{\alpha}_t=\prod_{i=0}^{t}(1-\beta_i)$ is the cumulative product of $(1-\beta_i)$ and $\{\beta_i\}$ follows a linear schedule. In our implementation, diffusion is performed with $T=100$ timesteps to balance sampling efficiency and mask fidelity.

\subsubsection{Prompt Encoding and Conditioning}
Each slice is assigned a binary prompt $y\in\{0,1\}$ indicating lesion absence/presence. A learnable embedding table maps $y$ to a vector $p(y)\in\mathbb{R}^{d_p}$ with $d_p=16$:
\begin{equation}
p(y) = \mathrm{Embed}(y).
\end{equation}
Conditioning is implemented by concatenation (rather than cross-attention), which is sufficient for a binary control signal and yields a lightweight denoiser. The denoiser input is:
\begin{equation}
\mathrm{inp} = [z_t \,;\, t/T \,;\, p(y)].
\end{equation}

\subsubsection{Denoising Network and Objective}
We train a neural network $\epsilon_\theta$ to predict the injected noise:
\begin{equation}
\hat{\epsilon} = \epsilon_\theta(z_t, t, p(y)).
\end{equation}
The diffusion model is trained with a mean squared error loss:
\begin{equation}
\mathcal{L}_{\text{diff}} = \mathbb{E}\left[\|\hat{\epsilon}-\epsilon\|_2^2\right].
\end{equation}
In practice, $\epsilon_\theta$ is implemented as an MLP with two hidden layers of width 1024 and SiLU activations, mapping from $\mathbb{R}^{D+1+d_p}$ to $\mathbb{R}^{D}$.

\subsubsection{Reverse Process and Mask Generation}
At inference, we initialize $z_T\sim\mathcal{N}(0,I)$ and iteratively apply the reverse denoising updates from $t=T-1$ down to $0$ to obtain $z_0$. The final mask logits are produced by the frozen VAE decoder:
\begin{equation}
\hat{L} = G_\phi(z_0),
\end{equation}
and the discrete synthetic mask is obtained by pixel-wise argmax:
\begin{equation}
\tilde{x}(u,v)=\arg\max_c \hat{L}_c(u,v).
\end{equation}
This yields anatomically coherent multi-class segmentation masks directly from noise, with optional lesion control via the prompt variable $y$.

\section{Results}

\subsection{Qualitative Mask Generation}

\begin{figure}[t]
    \centering
\caption*{\textbf{Top row: Real masks}}
\begin{subfigure}{0.24\linewidth}
    \centering
    \includegraphics[height=\linewidth, angle=90]{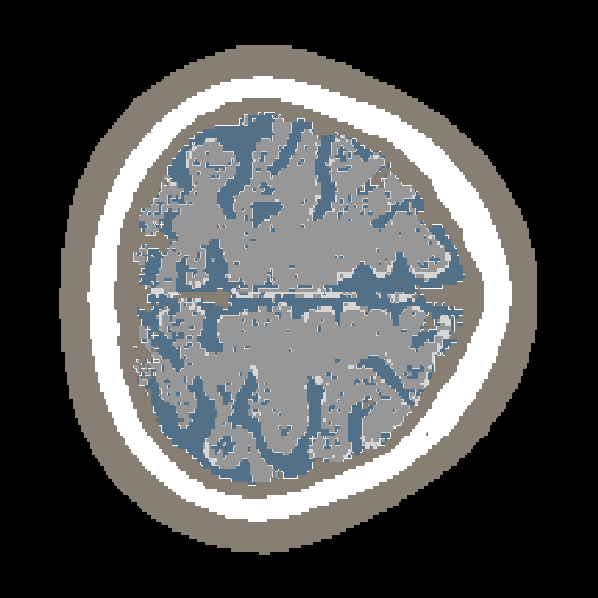}
\end{subfigure}
\hfill
\begin{subfigure}{0.24\linewidth}
    \centering
    \includegraphics[height=\linewidth, angle=90]{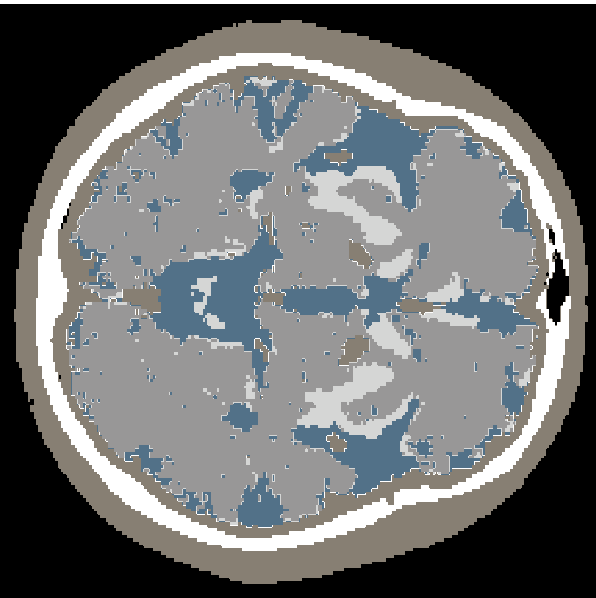}
\end{subfigure}
\hfill
\begin{subfigure}{0.24\linewidth}
    \centering
    \includegraphics[height=\linewidth, angle=90]{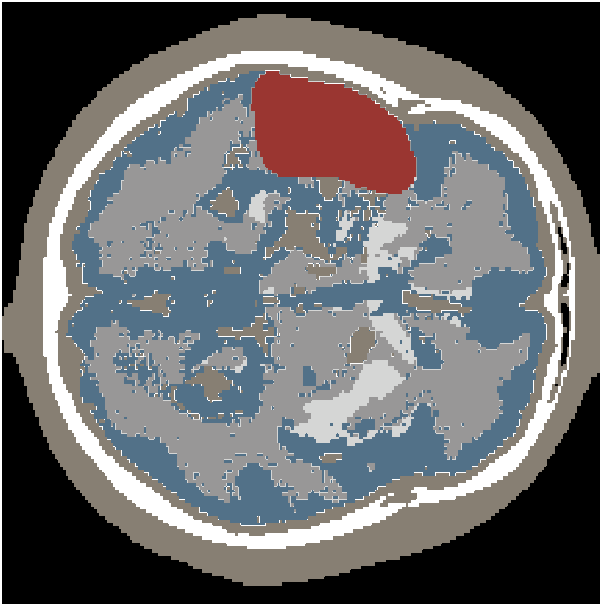}
\end{subfigure}
\hfill
\begin{subfigure}{0.24\linewidth}
    \centering
    \includegraphics[height=\linewidth, angle=90]{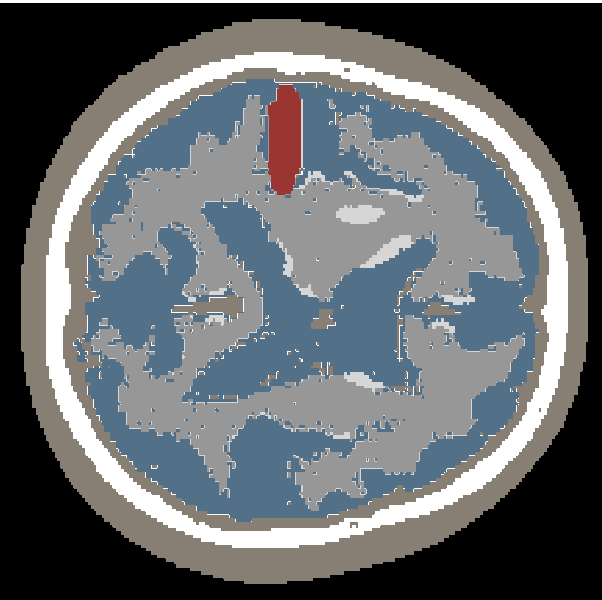}
\end{subfigure}

\vspace{0.6em}

\caption*{\textbf{Bottom row: Synthetic masks}}
\begin{subfigure}{0.24\linewidth}
    \centering
    \includegraphics[height=\linewidth, angle=90]{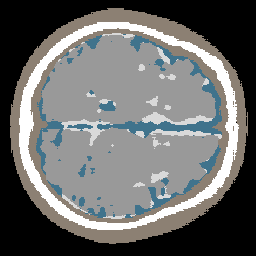}
\end{subfigure}
\hfill
\begin{subfigure}{0.24\linewidth}
    \centering
    \includegraphics[height=\linewidth, angle=90]{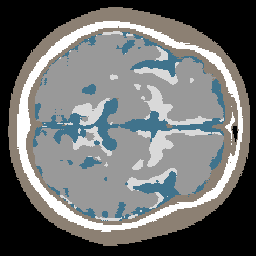}
\end{subfigure}
\hfill
\begin{subfigure}{0.24\linewidth}
    \centering
    \includegraphics[height=\linewidth, angle=90]{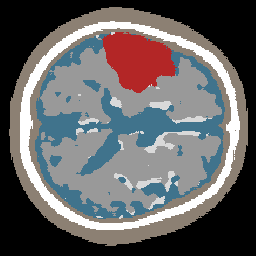}
\end{subfigure}
\hfill
\begin{subfigure}{0.24\linewidth}
    \centering
    \includegraphics[height=\linewidth, angle=90]{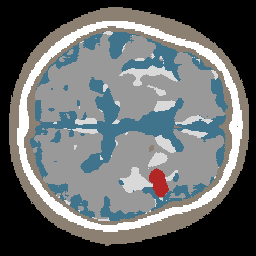}
\end{subfigure}

    \caption{Qualitative comparison between real (top row) and synthetic (bottom row) brain segmentation masks generated without lesion conditioning ($y=0$, left) and with lesion conditioning ($y=1$, right).}
    \label{fig:qualitative_results}
\end{figure}

Figure~\ref{fig:qualitative_results} shows representative examples of real and synthetic brain segmentation masks generated by the proposed framework. 
The top row corresponds to samples generated without lesion conditioning ($y=0$), while the bottom row shows samples generated with lesion conditioning ($y=1$). 
-
Across both settings, synthetic masks preserve global brain anatomy, clear tissue boundaries, and consistent class semantics comparable to real data. 
When lesion conditioning is enabled, ischemic infarct regions are generated without disrupting the overall anatomical structure.

\subsection{Class Distribution Analysis}

To further evaluate the statistical consistency between real and synthetic data, we analyze pixel-wise class distributions computed over the test set. 
The test set includes 473 lesion-free images and 136 images containing ischemic infarcts. 
For comparison, an equal number of synthetic masks was generated with ($y=1$) and without ($y=0$) lesion conditioning.

\begin{figure}[t]
    \centering
    \includegraphics[width=0.9\linewidth]{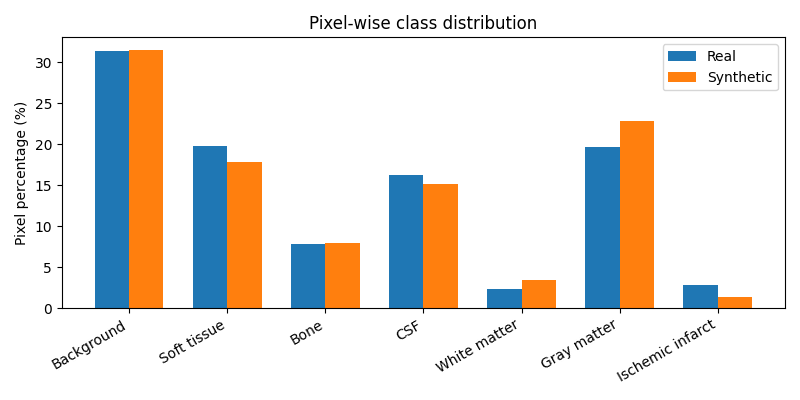}
    \caption{Pixel-wise class distribution comparison between real and synthetic brain segmentation masks evaluated on the test set.
    Distributions are shown for both lesion-free ($y=0$) and lesion-conditioned ($y=1$) samples.}
    \label{fig:classdist}
\end{figure}

Figure~\ref{fig:classdist} summarizes the pixel-wise class distribution across real and synthetic masks for both lesion-free and lesion-conditioned samples. 
Overall, the synthetic data closely match the real class proportions for background, soft tissue, bone, and cerebrospinal fluid, while preserving plausible relationships between gray and white matter. 
In lesion-conditioned samples, the relative prevalence of ischemic infarct pixels is comparable between real and synthetic masks, and minor redistributions among parenchymal tissues are consistent with infarcted regions replacing healthy tissue.

Although lesion conditioning is binary, the observed pixel distributions indicate natural variability in lesion extent across synthetic samples, suggesting that the latent diffusion process captures a considerabled spectrum of lesion burden without explicit constraints on lesion size or morphology.

In addition, we analyze the evolution of the Fréchet Inception Distance (FID) across training to select the final model checkpoint.

Table~\ref{tab:fid_epochs} reports FID values for representative epochs.
After the initial rapid decrease, the lowest FID is obtained at epoch 800, indicating improved distributional alignment between real and synthetic masks.
Therefore, the checkpoint at epoch 800 is used for all experiments presented in this work.

\begin{table}[t]
\centering
\caption{Fréchet Inception Distance (FID) evaluated at selected training epochs.
Lower values indicate higher similarity between real and synthetic segmentation masks.}
\label{tab:fid_epochs}
\begin{tabular}{l c}
\hline
\textbf{Checkpoint} & \textbf{FID} \\
\hline
Epoch 100 & 64.43 \\
Epoch 400 & 63.91 \\
Epoch 800 & \textbf{61.88} \\
\hline
\end{tabular}
\end{table}

\section{Discussion}
This work addresses the problem of unconditional generation of multi-class brain segmentation masks, with a focus on preserving anatomical structure and tissue semantics. 
By decoupling anatomical representation learning from stochastic generation, the proposed framework enables stable synthesis of anatomically plausible masks directly from noise, without requiring conditioning images or spatial guidance at inference time.

A central strength of the approach lies in constraining diffusion to the latent space of a variational autoencoder trained exclusively on segmentation masks. 
This design enforces global anatomical consistency and prevents the generation of structurally implausible configurations commonly observed in pixel-space generative models for discrete labels.

The inclusion of a binary lesion prompt provides limited yet effective semantic control over infarct presence. 
Despite its simplicity, this mechanism allows pathological regions to be introduced while maintaining overall anatomical coherence, and naturally yields variability in lesion extent across samples.

Several limitations should be acknowledged. 
The framework operates on 2D axial slices and does not enforce volumetric consistency, and lesion control is restricted to binary presence or absence. 
Furthermore, evaluation focuses on qualitative and distributional analyses rather than downstream task performance. 
Addressing these limitations constitutes an important direction for future work.

Despite these constraints, the proposed method offers a practical and scalable solution for generating anatomically valid brain segmentation masks in data-scarce settings, and provides a useful structural prior for data augmentation and image synthesis applications.

\section{Conclusions}

We introduced an anatomy-preserving latent diffusion framework for the unconditional generation of multi-class brain segmentation masks from NCCT data. 
By performing diffusion in the latent space of a variational autoencoder trained on segmentation masks, the proposed approach constrains generation to anatomically plausible configurations while enabling efficient sampling from pure noise.

The results demonstrate that the model generates diverse and realistic brain masks that preserve global anatomical structure and tissue organization, with optional semantic control over lesion presence. 
Compared to pixel-space generative models, the proposed framework avoids structural artifacts and label inconsistencies, making it well suited for discrete anatomical representations.

Future work will explore extensions to volumetric 3D modeling, finer-grained lesion control, and quantitative evaluation in downstream tasks such as segmentation model training and clinical simulation.

\section{Data and Code Availability}

The code used to generate the synthetic segmentation masks is available at
\href{https://github.com/lucixia3/synthetic_masks_generation.git}{https://github.com/lucixia3/synthetic\_masks\_generation.git},
and the resulting dataset is available at
\href{https://huggingface.co/datasets/SEARCH-IHI/IISM_brain}{https://huggingface.co/datasets/SEARCH-IHI/IISM\_brain}.

\section{Acknowledgment}
This work is part of the European project SEARCH, which is supported by the Innovative Health Initiative Joint Undertaking (IHI JU) under grant agreement No. 101172997. The JU receives support from the European Union’s Horizon Europe research and innovation programme and COCIR, EFPIA, Europa Bio, MedTech Europe, Vaccines Europe, Medical Values GmbH, Corsano Health BV, Syntheticus AG, Maggioli SpA, Motilent Ltd, Ubitech Ltd, Hemex Benelux, Hellenic Healthcare Group, German Oncology Center, Byte Solutions Unlimited, AdaptIT GmbH. Views and opinions expressed are however those of the author(s) only and do not necessarily reflect those of the aforementioned parties. Neither of the aforementioned parties can be held responsible for them.

\vspace{12pt}
\color{red}

\end{document}